\documentclass[pra,aps,twocolumn,nopacs,superscriptaddress,nofootinbib,longbibliography]{revtex4-1} 

\usepackage{graphicx}
\usepackage{psfrag}
\usepackage{epsfig}
\usepackage{pst-all}
\usepackage{gensymb}
\usepackage{bm}
\usepackage{bbm}
\usepackage{braket}
\usepackage{dcolumn}
\usepackage{amsmath}
\usepackage{amssymb}
\usepackage{amsfonts}\usepackage{epsfig}
\usepackage{mathrsfs}
\usepackage{mathtools}
\usepackage{lmodern}
\usepackage{indentfirst}
\usepackage{color}
\usepackage[T1]{fontenc}
\usepackage{comment}
\usepackage{xcolor}
\usepackage[colorlinks,linkcolor=blue,citecolor=blue,urlcolor=blue,hyperindex,driverfallback=dvipdfm]{hyperref}
\usepackage{soul}

\def\ii{{\rm i}}  \def\ee{{\rm e}}
  
\def\rb{{\bf r}}  \def\Rb{{\bf R}}    \def\vb{{\bf v}}
    \def\zz{\hat{\bf z}}

\def\kb{{\bf k}}

\def\me{m_{\rm e}}  

\begin{document} 

\def\bibsection{\section*{\refname}} 

\title{Free-Electron Shaping Using Quantum Light
}


\author{Valerio~Di~Giulio}
\affiliation{ICFO-Institut de Ciencies Fotoniques, The Barcelona Institute of Science and Technology, 08860 Castelldefels (Barcelona), Spain}
\author{F.~Javier~Garc\'{\i}a~de~Abajo}
\email{javier.garciadeabajo@nanophotonics.es}
\affiliation{ICFO-Institut de Ciencies Fotoniques, The Barcelona Institute of Science and Technology, 08860 Castelldefels (Barcelona), Spain}
\affiliation{ICREA-Instituci\'o Catalana de Recerca i Estudis Avan\c{c}ats, Passeig Llu\'{\i}s Companys 23, 08010 Barcelona, Spain}




\begin{abstract}
Controlling the wave function of free electrons is important to improve the spatial resolution of electron microscopes, the efficiency of electron interaction with sample modes of interest, and our ability to probe ultrafast materials dynamics at the nanoscale. In this context, attosecond electron compression has been recently demonstrated through interaction with the near fields created by scattering of ultrashort laser pulses at nanostructures followed by free electron propagation. Here, we show that control over electron pulse shaping, compression, and statistics can be improved by replacing coherent laser excitation by interaction with quantum light. We find that compression is accelerated for fixed optical intensity by using phase-squeezed light, while amplitude squeezing produces ultrashort double-pulse profiles. The generated electron pulses exhibit periodic revivals in complete analogy to the optical Talbot effect. We further reveal that the coherences created in a sample by interaction with the modulated electron are strongly dependent on the statistics of the modulating light, while the diagonal part of the sample density matrix reduces to a Poissonian distribution regardless of the type of light used to shape the electron. The present study opens a new direction toward the generation of free electron pulses with additional control over duration, shape, and statistics, which directly affect their interaction with a sample.
\end{abstract}

\maketitle 
\date{\today} 
\tableofcontents 
\setcounter{equation}{0} 


\section{Introduction}

The exploration of ultrafast phenomena generally relies on the use of short probe pulses, such as those provided by femtosecond visible-infrared lasers and attosecond x-ray sources \cite{PTB01,CK07,KI09}. Electrons can potentially reach much shorter durations than light for typical beam energies in the $10^2$-$10^5\,$eV range, as they are characterized by oscillation periods of 20-0.02\,as. Electron pulse compression is also capital for free-electron lasers \cite{MT10}, relying on the $\propto N^2$ superradiance emission produced by $N$ electrons when acting as a single point charge. With applications such as imaging, spectroscopy, and light generation in view, strong interest has arisen in manipulating the free electron density matrix using light.

Triggered by the advent of the so-called photon-induced near-field electron microscopy (PINEM) \cite{BFZ09}, a long series of experimental \cite{BFZ09,KGK14,PLQ15,FES15,paper282,EFS16,KSE16,RB16,VFZ16,KML17,KES17,FBR17,PRY17,paper306,paper311,MB18,MB18_2,paper332,DNS20,KLS20,WDS20} and theoretical \cite{paper151,PLZ10,PZ12,B17_2,paper272,paper312,K19,RML19,paper347} studies have demonstrated that interaction with the optical near fields scattered from illuminated nanostructures provides an efficient way to manipulate the temporal and spatial distribution of free electrons. In PINEM, electron and light pulses are made to interact in the presence of a sample, giving rise to multiple photon exchanges between the optical field and the electron, and leading to comb-like energy spectra characterized by sidebands that are associated with different numbers of exchanged photons and separated from the incident electron energy by a multiple of the photon energy. Recent experiments have measured hundreds of such sidebands produced through suitable combinations of sample geometry and illumination conditions \cite{DNS20,KLS20}. Additionally, electron pulse compression has been observed by free propagation of PINEM-modulated electrons over a sufficiently long distance \cite{KML17,PRY17,MB18_2,MB18}. The electron transforms into a series of pulses with duration down to the attosecond regime \cite{KML17,PRY17}, which can be made even smaller by increasing the strength of the PINEM light \cite{B17_2}.

While this type of electron-light interaction affects only the longitudinal part of the electron wave function, lateral control can be achieved either by the use of electron phase masks \cite{VTS10,VGC14,VBM18,SLR19} or through modulating the optical field with a transverse spatial resolution limited by the light wavelength, and more generally, by the polariton wavelength when relying on the excitation of optical modes in material surfaces. By analogy to elastic electron diffraction by light gratings in free space (the Kapitza-Dirac effect \cite{KD1933,FAB01,FB02}), which has been shown to also enable the formation of vortex beams \cite{HSB15}, surface-plasmon standing waves can produce intense inelastic electron diffraction \cite{paper272}, as confirmed by the observation of discrete electron beam deflection upon absorption or emission of a given number of photons reflected from a thin metal plate \cite{paper311}. Similarly, optical near fields can transfer orbital angular momentum \cite{paper312}, also demonstrated through the synthesis and observation of vortex electron beams produced by inelastic interaction with chiral near fields \cite{paper332}. As a practical application of these phenomena, lateral phase imprinting on electron beams through optical fields has been recently proposed to provide a viable approach to aberration correction and lateral electron beam profiling \cite{paper351}.

By sweeping the photon energy of the light used for PINEM interaction, the near field experienced by the electrons undergoes amplitude modulations that map the optical response of the sample. This strategy has been proposed as a form of spectrally-resolved microscopy that can combine the subnanometer spatial focusing of electron beams \cite{BDK02} with an excellent energy resolution limited by the spectral width of the light source \cite{H99,paper114}. A first demonstration of this possibility has enabled spatial mapping of plasmons in silver nanowires with $\sim20\,$meV energy resolution without any need for electron monochromators \cite{paper306}, a result that is rivalling the energy resolution achieved through state-of-the art electron energy-loss spectroscopy \cite{KLD14}.

The above studies rely on coherent light, such as that generated by laser sources, while an extension to quantum optical fields has been recently predicted to introduce quantum effects in the electron spectra \cite{paper339}. Quantum light thus presents an opportunity to further manipulate the electron wave function in applications such as pulse compression and modulation of the electron statistics.

Here, we show that a wide range of electron statistics can be reached through interaction of free electrons with quantum light. Besides changing the focusing properties of the optically-modulated electrons, this interaction reveals a strong dependence of the electron density matrix on the statistics of the light field, which can be observed in a self-interference configuration setup. Specifically, we show that interaction with phase-squeezed and minimum-phase-uncertainty light sources produce faster compression of the electron, while amplitude-squeezed light gives rise ultrashort double-pulse electron profiles. Additionally, we find that the interaction of the modulated electron with a target produces a Poissonian distribution of sample excitations with off-diagonal coherences that are strongly dependent on the statistics of the light used to modulate the electron. Besides the fundamental interest of this wealth of phenomena, we envision applications in the control of electron compression and in the generation of light with nontrivial statistics.

\section{Electron density matrix produced upon PINEM interaction}
\label{quantumPINEM}

\subsection{The quantum PINEM interaction}
\label{quantumPINEMinteraction}

Free electron-light interaction has been extensively studied under the assumption of classical illumination \cite{paper151,PLZ10}. An extension to describe the quantum evolution of the joint electron-light state has been recently presented \cite{paper339}, which we use here to investigate the modification produced in the electron density profile following propagation after PINEM interaction with nonclassical light. We first provide a succinct summary of this quantum formalism.

We consider the sample response to be dominated by a single bosonic optical mode oscillating at frequency $\omega_0$ and characterized by an electric-field distribution $\vec{\mathcal{E}}_0(\rb)$ defined as either a normal \cite{GL91} or a quasi-normal \cite{FHD19} bosonic mode. In addition, we assume that the electron always consists of a superposition of states with relativistic momentum and energy tightly focused around $\hbar\kb_0$ and $E_0$ (i.e., having small uncertainties compared with $\hbar\omega_0/v$ and $\hbar\omega_0$, respectively, where $v$ is the electron velocity). Also, we ignore nonunitary elements in the dynamics by considering that the electron-light interaction happens on a fast time scale compared with the decay of the bosonic mode. These assumptions allow us to linearize the electron kinetic energy operator (nonrecoil approximation). Starting from the Dirac equation \cite{S1994} and following an approach inspired by quantum optics methods \cite{SZ97} with an electromagnetic gauge in which the scalar potential is zero, the effective Hamiltonian of the system can be approximated by the noninteraction and interaction pieces \cite{paper339}
\begin{subequations}
\begin{align}
\hat{\mathcal{H}}_0&=\hbar \omega_0 a^\dagger a + E_0  -\hbar \vb \cdot (\ii \nabla + \kb_0), \\
\hat{\mathcal{H}}_1&= -\ii  (e \vb/\omega_0) \cdot \left[\vec{\mathcal{E}}_0(\rb) a - \vec{\mathcal{E}}_0^*(\rb) a^\dagger\right],
\end{align}
\label{H0H1}
\end{subequations}
respectively, where $a$ and $a^\dagger$ are annihilation and creation operators of the bosonic optical mode, and $\vb =\hbar\kb_0/E_0=v\zz$ is the electron velocity vector, taken to be along $\zz$. We remark that the aforementioned QED model accurately reproduces the electron-field dynamics when spin-flips, ponderomotive forces, and electron recoil can be safely disregarded. However, in situations departing from these conditions, the full minimal-coupling Hamiltonian has to be considered, and thus, numerical integration provides a more suitable method to explore the resulting physics \cite{T17,T18,T20}. We can then write the solution for the electron-optical mode wave function as a sum of energy sidebands, each of them describing the amplitude associated with a net exchange of $\ell$ quanta with the optical mode ($\ell>0$ for electron energy gain and $\ell<0$ for loss). More precisely, we have (see Ref.\ \cite{paper339} and Appendix\ \ref{appendixA})
\begin{align}
|\psi(\rb,t)\rangle=&\psi_{\rm inc}(\rb,t)\!\!\sum_{\ell=-\infty}^\infty\sum_{n=0}^\infty\ee^{\ii \omega_0[\ell(z/v-t)-n t]}f_\ell^n(\rb)|n\rangle, \label{eq:solution}
\end{align}
where $\rb$ denotes the electron coordinate, $|n\rangle$ runs over Fock states of the optical field, $\psi_{\rm inc}(\rb,t)$ is the incident electron wave function, and the amplitude coefficients admit the closed-form expression
\begin{align}
f_\ell^n=&\ee^{\ii (\chi+\ell {\rm arg}\{-\beta_0\})}\, \alpha_{n+\ell}\, F_\ell^n \label{eq:intcoeff}\\
F_\ell^n=&|\beta_0|^{\ell}\ee^{-|\beta_0|^2/2}\sqrt{(n+\ell)!n!}
\!\!\!\!\!\!\!\!\!\! \sum_{n'={\rm max}\{0,-\ell\}}^n \!\!\!\!\!\!\!\!\!\! \frac{(-|\beta_0|^2)^{n'}}{n'!(\ell+n')!(n-n')!}, \nonumber 
\end{align}
with
\begin{align}
\beta_0(\Rb,z)=\frac{e}{\hbar\omega_0}\int_{-\infty}^z dz'\,\mathcal{E}_{0,z}(\Rb,z')\ee^{-\ii \omega_0 z'/v}
\nonumber
\end{align}
acting as a single-mode coupling coefficient and $\chi=(-e/\hbar\omega_0)\int_{-\infty}^z dz'\,{\rm Im}\{\beta_0^*(\Rb,z')\mathcal{E}_{0,z}(\Rb,z')\ee^{-\ii \omega_0 z'/v}\}$ representing a global phase that is irrelevant in the present study. A dependence on lateral coordinates $\Rb=(x,y)$ is imprinted by the spatial distribution of the optical mode field. In the initial state (i.e., before quanta exchanges), only $\ell=0$ terms are present, so we can write $f^n_\ell(z\rightarrow -\infty)=\delta_{\ell0}\alpha_n$, where the amplitudes $\alpha_n$ define the starting optical boson field, which must satisfy the normalization condition
\begin{align}
\sum_n|\alpha_n|^2=1.
\label{sumalphan}
\end{align}
Interestingly, the number of excitations $n'=n+\ell$ is conserved along the temporal evolution of the system \cite{paper339}, thus allowing us to propagate each initial $n'$ component separately and multiply it by the initial boson amplitude $\alpha_{n+\ell}$ when writing Eq.\ (\ref{eq:intcoeff}). Because the expansion coefficients defined in this equation are obtained from the evolution operator \cite{paper339}, they satisfy the normalization condition $\sum_{\ell n} |f_\ell^n|^2=\sum_{\ell n'} |\alpha_{n'}F_\ell^{n'-\ell}|^2=1$ for any optical field, which leads to the condition
\begin{align}
\sum_\ell(F_\ell^{n-\ell})^2=1
\label{sumFln}
\end{align}
satisfied for any $n$.

Electron propagation prior to interaction is described through the linearized Hamiltonian $\hat{\mathcal{H}}_0$, which essentially assumes that the electron beam is well collimated and energy dispersion is negligible in the PINEM interaction region, such that we can write
\begin{align}
\psi_{\rm inc}(\rb,t)=\ee^{\ii\kb_0\cdot \rb-\ii E_0t/\hbar}\phi(\rb-\vb t),
\nonumber
\end{align}
where $\phi$ is a slowly varying function of relative position $\rb-\vb t$. Importantly, Eq.\ (\ref{eq:intcoeff}) prescribes that the evolution of the electron-boson system is uniquely determined by the nondimensional coupling parameter $\beta_0$ in combination with the amplitudes $\alpha_n$ defining the initial optical wave function. In what follows, we assume no dependence on $\Rb$ (see below) and set $\beta_0\equiv\beta_0(z\rightarrow\infty)$ because we are interested in studying free-electron propagation after PINEM interaction has taken place, even though this dependence plays a fundamental role in the observed transfer of orbital angular momentum between photons and electrons \cite{paper332}, and in addition, it could be useful to correct electron beam aberrations \cite{paper351}. Nevertheless, the coefficients of the quantum light state in Eq.\ (\ref{eq:intcoeff}) could provide an additional knob to further intertwine longitudinal and transverse electron degrees of freedom beyond what is possible using classical light. Additionally, they could affect the maximum achievable probability associated with specific PINEM sidebands, as well as the dependence on pulse duration, which also deserve further study.

\begin{figure*}
\centering{\includegraphics[width=1.0\textwidth]{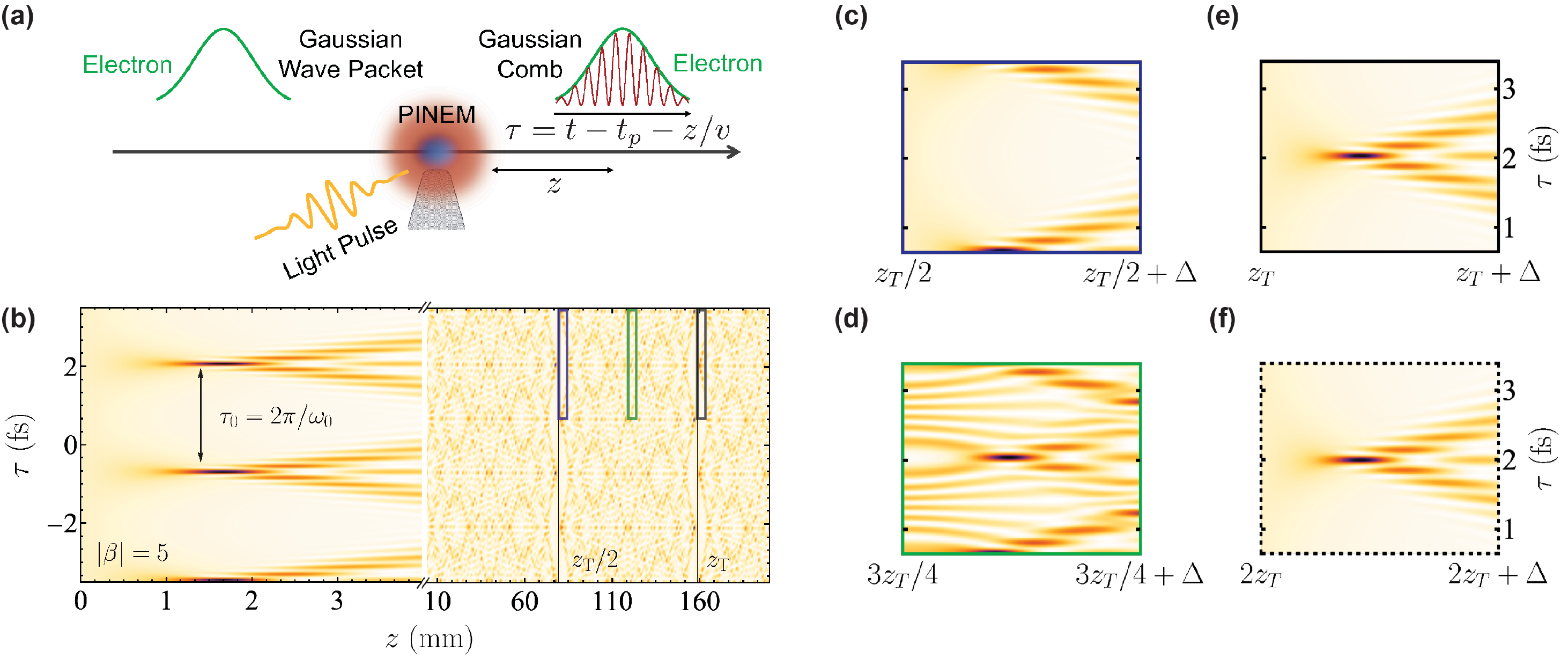}}
\caption{{\bf Talbot effect and electron compression with classical light.} (a) An electron Gaussian wave packet (green) is transformed through PINEM interaction followed by propagation along a distance $z$ into a substantially modified electron density profile in the propagation-distance-shifted time $\tau=t-z/v$ due to superposition of different energy components. (b) Electron density profile (vertical $\tau$ coordinate) as a function of propagation distance $z$ (horizontal axis) after PINEM interaction with coherent light. We consider $100$\,keV electrons, a photon energy $\hbar\omega_0=1.5$\,eV, and a coupling coefficient $|\beta|=5$. Trains of compressed electron pulses are periodically observed at discrete multiple values of the Talbot propagation distance $z_T$. (c-e) Details of the $\tau$-$z$ map in (b) corresponding to the color-matched square regions of $z$ width $\Delta=4\,$mm. (f) Same as (e), but for z near $2z_{\rm T}$.}
\label{Fig1}
\end{figure*}

\subsection{Effect of free propagation}

Our purpose is to investigate the electron characteristics after free propagation over a macroscopic distance of several mm from the PINEM interaction region [see Fig.\ \ref{Fig1}(a)]. We identify in Eq.\ (\ref{eq:solution}) a propagation phase $\ee^{\ii k_\ell z}$ associated with each $\ell$ sideband, in which the electron wave vector is replaced by its linearized nonrecoil version $k_\ell\approx k_0 + \ell \omega_0/v$. While this approximation does accurately describe propagation over the relatively small extension of the PINEM interaction region, the exact expression
\begin{align}
k_\ell&=\hbar^{-1}\!\sqrt{E_\ell^2/c^2-\me^2 c^2 } \label{klexpan}\\
&\approx k_0+\ell\omega_0/v-2\pi\ell^2/z_T+\cdots,
\nonumber
\end{align}
needs to be used to deal with arbitrarily long propagation distances $z$, where the second-order correction, characterized by a distance
\begin{align}
z_T=4\pi\,\me v^3\gamma^3/\hbar\omega_0^2
\label{zT}
\end{align}
(e.g., $z_T\approx159\,$mm for $\hbar\omega_0=1.5\,$eV and 100\,keV electrons), is sufficiently accurate under the conditions here considered, giving rise to numerical results that are indistinguishable from the full expression in the examples shown below.

Our purpose is to study electron propagating and dismiss any entanglement with the PINEM optical field. We thus consider the electron density matrix, obtained from the pure-joint-state density matrix $|\psi(z,t)\rangle\langle\psi(z',t)|$ by tracing out the optical degrees of freedom: 
\begin{align}
\rho(z,z',t)=\sum_{n=0}^\infty \psi_n(z,t)\psi_n^*(z',t), \label{rho1}
\end{align}
with
\begin{align}
\psi_n(z,t)=\phi(z-vt) \sum_{\ell=-\infty}^\infty \alpha_{n+\ell}\, F_\ell^n \ee^{\ii k_\ell z-\ii\ell\omega_0(t-t_p)}, \nonumber
\end{align}
where the phase of $\beta_0$ enters only through a time shift $t_p={\rm arg}\{-\beta_0\}/\omega_0$. We remark here that the mathematical operation of tracing out the degrees of freedom associated with the photonic mode to obtain a density matrix for the electron subsystem is physically justified by the fact that this operation ensures the correct measurement statistics if one only needs to measure electron properties (i.e., without performing any measurement on the rest of the system)\cite{NC04}.

We note that diffraction effects involving the transverse evolution of the wave function are disregarded. Under attainable experimental conditions, an initial 100\,keV electron beam with $\varphi\sim50\,\mu$rad divergence, focused to a $2/k_0\varphi\sim25\,$nm spot over the PINEM interaction region, becomes just a factor $\sim2$ wider after free propagation over a distance $z\sim1\,$mm due to diffraction. In addition, the results here presented are valid under the assumption that $\phi(z-vt)$ involves a sufficiently narrow wave vector decomposition to neglect corrections beyond the linear energy dependence of the wave vector during the propagation distances under consideration, so $\phi$ enters the electron density matrix just as a broad envelope factor. However, we note that these assumptions may break in scenarios involving slow electrons ($E_0\lesssim10^2$ eV) or very strong electron-field coupling, in which the ponderomotive force can lead to a non-negligible beam spreading after interaction with the sample \cite{T20}.

\subsection{Talbot effect and periodicity of the density matrix}

Retaining just up to $\ell^2$ corrections in Eq.\ (\ref{klexpan}) for $k_\ell$ and considering relative positions $|z-z'|\ll z_T$, we can recast the electron density matrix (Eq.\ (\ref{rho1})) as
\begin{align}
\rho(z,z',t)=\ee^{\ii k_0(z-z')}\phi(z-vt)\phi^*(z'-vt)\tilde\rho(z,\tau,\tau'), \nonumber
\end{align}
where
\begin{align}
\tilde\rho(z,\tau,\tau')=\sum_{n\ell\ell'} &\alpha_{n+\ell}\alpha_{n+\ell'}^* \; F_\ell^n F_{\ell'}^n \label{rhotilde}\\ &\times\ee^{2\pi\ii\left[({\ell'}^2-\ell^2)z/z_T+(\ell'\tau'-\ell\tau)/\tau_0\right]}, \nonumber
\end{align}
$\tau=t-t_p-z/v$, and $\tau'=t-t_p-z'/v$. Disregarding the trivial phase propagation factor $\ee^{\ii k_0(z-z')}$ and the slowly varying envelope introduced by $\phi$, the density matrix is periodic in both of the time-shifted coordinates $\tau$ and $\tau'$ with the same period as the light optical cycle $\tau_0=2\pi/\omega_0$. Additionally, we find that $\tilde\rho(z,\tau,\tau')$ portrays a periodic pattern as a function of propagation distance $z$ similar to the Talbot effect \cite{T1836,R1881,LS1971_2,L1988,NT93}, with a period given by $z_T$ (Eq.\ (\ref{zT})).

To illustrate this effect, we plot in Fig.\ \ref{Fig1}(b) the diagonal elements $\rho(z,z,t)=\sum_{n=0}^\infty \left|\psi_n(z,t)\right|^2$ normalized to the envelope density $|\phi(z-vt)|^2$ for coherent light illumination, which represent the scaled electron density profile as a function of time and propagation distance $z$ from the PINEM interaction region, calculated in the high-fluence classical limit (see below). Incidentally, off-diagonal elements are also considered and represented below in Fig.\ \ref{Fig4}. The plot clearly reveals a train of temporally focused electron pulses at $z\sim1.5\,$mm, followed by a series of focusing revivals at intervals of $z_T\approx159\,$mm and accompanied by temporally shifted revivals at fractional values of the Talbot distance $z_T$ \cite{BK96}.

\section{Electron pulse compression with different optical mode statistics}

Before analyzing the effect of light statistics in the evolution of the electron after PINEM interaction, we remark that the previous formalism is only valid for pure initial optical states, whose density matrix is given by $\sum_{nn'}\alpha_n\alpha_{n'}^*|n\rangle\langle n'|$. In contrast, for a perfect mixture (i.e., an initial optical density matrix $\sum_n|\alpha_n|^2|n\rangle\langle n|$ with no coherences), the outcome of interaction and propagation has to be separately calculated for each Fock state $|n\rangle$ and then averaged incoherently. Using the normalization conditions of Eqs.\ (\ref{sumalphan}) and (\ref{sumFln}), we find an electron density matrix $\tilde{\rho}(z,\tau,\tau')=1$, which is not altered due to interference between different energy components after PINEM interaction. We note that a well-defined optical Fock state belongs to this category and thus does not produce changes in the electron density matrix either.

\begin{figure}
\centering{\includegraphics[width=0.45\textwidth]{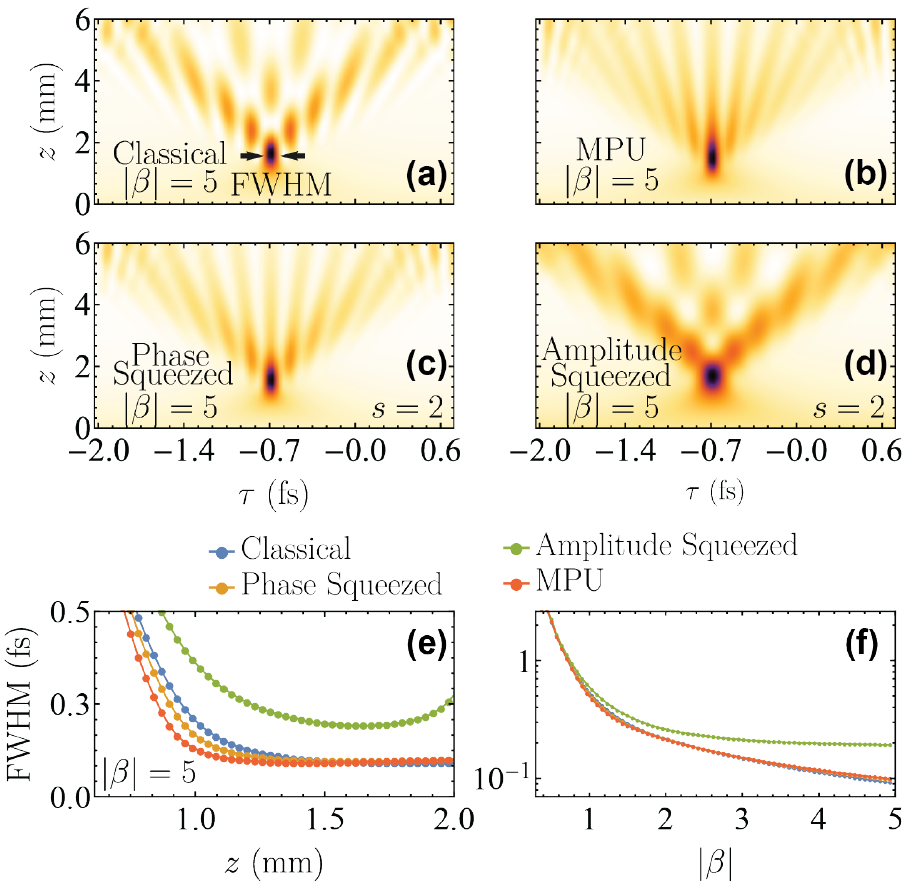}}
\caption{{\bf Electron compression using squeezed light.} (a-d) Evolution of the electron density profile following PINEM interaction with (a) classical, (b) MPU, (c) phase-squeezed, and (d) amplitude-squeezed light using a single-mode coupling coefficient $|\beta_0|=0.2$ and average population $\bar{n}=625$ (i.e., $|\beta|=\sqrt{\bar{n}}|\beta_0|=5$). (e) FWHM [see panel (a)] of the compressed electron density in (a-d) as a function of propagation distance $z$. (f) Minimum in the FWHM along the curves in (e) as a function of coupling coefficient $|\beta|$ (varying $|\beta_0|$ and keeping $\bar{n}=625$). We consider $100$\,keV electrons and a $1.5$\,eV photon energy.}
\label{Fig2}
\end{figure}

\subsection{High-fluence and classical limits}

Electron coupling to a single optical mode is generally weak and therefore characterized by a small coupling coefficient $|\beta_0|\ll1$ (e.g., we set $|\beta_0|=0.2$ here, as a feasible value for coupling to Mie and plasmon modes in nanoparticles \cite{paper339}). Still, a strong PINEM effect can be produced with a high average number of photons $\bar{n}=\sum_n n|\alpha_n|^2$, while only sidebands $|\ell|\ll\bar{n}$ can then be efficiently populated. In this limit, using the Stirling formula to approximate the factorials containing $n$ in Eq.\ (\ref{eq:intcoeff}), we find (see Appendix\ \ref{appendixB})
\begin{align}
F_\ell^n\approx J_\ell(2\sqrt{n}|\beta_0|). \label{largen}
\end{align}
Additionally, if the optical mode is prepared in a coherent state (e.g., by exciting it with laser light), its population follows a Poissonian distribution $|\alpha_n|^2=\ee^{-\bar{n}}\,\bar{n}^{n}/n!$,  which approaches a normal distribution \cite{F1968} $|\alpha_n|^2\approx\ee^{-(n-\bar{n})^2/2\bar{n}}/\sqrt{2\pi \bar{n}}$ for $\bar{n}\gg1$. Introducing this expression in Eq.\ (\ref{rhotilde}), approximating $n\approx\bar{n}$ in Eq.\ (\ref{largen}), and using the normalization condition $\sum_n|\alpha_n|^2=1$, we can write the density matrix in the high-fluence classical limit as
\begin{align}
\tilde\rho(z,\tau,\tau')\approx	\psi_{\rm cl}(z,\tau)\psi_{\rm cl}^*(z,\tau'), \nonumber
\end{align}
where
\begin{align}
\psi_{\rm cl}(z,\tau)=\sum_{\ell} &J_\ell(2|\beta|)\ee^{-2\pi\ii(\ell^2z/z_T+\ell\tau/\tau_0)} \nonumber
\end{align}
and
\begin{align}
\beta=\sqrt{\bar{n}}\beta_0 \label{beta}
\end{align}
is the effective coupling coefficient, which is proportional to the light intensity used to excite the optical mode. This result is consistent with previous theoretical \cite{FES15,B17_2} and experimental \cite{PRY17,MB18_2} studies of free propagation after high-fluence classical PINEM interaction. Electron compression and Talbot revivals in this limit are shown in Fig.\ \ref{Fig1}(b) for coherent illumination with $|\beta_0|=0.2$ and $\beta=5$, while a zoom of the focal region is presented in Fig.\ \ref{Fig2}(a).

Interestingly, for any population of the optical mode that is smooth and strongly peaked around $\bar{n}\gg1$, we can approximate $\alpha_{n+\ell}\approx\alpha_{n}$ for $|\ell|\ll n$, so the wave function completely separates into light and electron components in Eq.\ (\ref{eq:solution}), which becomes $|\psi(\rb,t)\rangle\approx\left\{\sum_{n=0}^\infty\alpha_{n}\ee^{-\ii n\omega_0t}|n\rangle\right\}\times\left\{\psi_{\rm inc}(\rb,t)\sum_{\ell=-\infty}^\infty\ee^{\ii (\chi+\ell {\rm arg}\{-\beta\})}\,J_\ell(2|\beta|)\,\ee^{\ii\ell\omega_0(z/v-t)}\right\}$, in agreement with a well-known expression for PINEM with classical light \cite{paper311}.

\subsection{Coherent squeezed light}

We now explore squeezed light as an experimentally feasible alternative to classical laser light to excite the PINEM optical mode. Single-mode coherent squeezed states $D(g)S(\zeta)|0\rangle$ are defined by applying the displacement and squeezing operators, $D(g)=\exp(ga^\dagger-g^*a)$ and $S(\zeta)=\exp\left[(\zeta^*aa-\zeta a^{\dagger}a^{\dagger})/2\right]$, to the optical vacuum \cite{LK1987}. Writing the squeezing parameter as $\zeta=\ee^{\ii \theta}s$, one can express the expansion coefficients of these states in the number basis representation as
\begin{align}
\alpha_n=\frac{\left(\xi/2\right)^{n/2}}{\sqrt{n!\cosh{s}}}\; \ee^{-(|g|^2+g^{* 2}\xi})/2 H_n\left[\frac{g+g^*\xi}{\sqrt{2\xi}}\right],\nonumber
\end{align}
where $\xi=\ee^{\ii\theta}\tanh{s}$ and $H_n$ is the Hermite polynomial of order $n$. These coefficients reduce to those of a coherent state for $s=0$. The average photon number is given by $\bar{n}=|g|^2+\sinh^2 s$, while $\alpha_n$ depends on the phases of $g$ and $\zeta$ through the combination $\varphi={\rm arg}\{g\}-\theta/2$. In particular, the variance takes minimum and maximum values for $\varphi=0$ and $\pi$, corresponding to amplitude- and phase-squeezed states, respectively \cite{LK1987}.

We consider the two extreme possibilities of PINEM interaction with purely phase- and amplitude-squeezed light in Fig.\ \ref{Fig2}(c,d), where we plot the density profile $\rho(z,z,t)=\tilde{\rho}(z,\tau,\tau)$ as a function of propagation distance $z$ for fixed coupling strength [$|\beta|=5$, obtained with $\bar{n}=625$ and $|\beta_0|=0.2$, see Eq.\ (\ref{beta})]. Electron focusing takes place at a similar propagation distance $z\sim2\,$mm for all light statistics under consideration. When the illumination has classical [Fig.\ \ref{Fig2}(a)] or amplitude-squeezed [Fig.\ \ref{Fig2}(d)] statistics, the density shows oscillations as a function of relative time $\tau$ before focusing. These oscillations disappear with phase-squeezed light [Fig.\ \ref{Fig2}(c)]. Additionally, the latter produces a focal spot spanning a larger interval of propagation distances $z$ and emerging at a shorter value of $z$ in comparison with classical light [Fig.\ \ref{Fig2}(e)]. The behavior with amplitude-squeezed light is the opposite, and in particular, the minimum full width at half maximum (FWHM) of the focal spot is approximately twice larger than the result obtained with phase-squeezed or classical light. As already discussed for classical light \cite{B17_2}, the degree of compression increases with increasing coupling $|\beta|$ [Fig.\ \ref{Fig2}(f)].

Incidentally, upon visual inspection of the $z$-$\tau$ pattern for coherent-state illumination in Fig.\ \ref{Fig2}(a), smoothing along $z$ would lead to vertical elongation of the density features, similar to those obtained using phase-squeezed light [Fig.\ \ref{Fig2}(b)]; in contrast, smoothing along $\tau$ would produce a pattern more similar to that of amplitude-squeezed illumination [Fig.\ \ref{Fig2}(d)]. This is consistent with the intuitive picture that phase-squeezing should generate sharper features in the wave function snapshots (i.e., narrower peaks as a function of $\tau$, accompanied by broadening along $z$ in order to preserve the total electron probability); conversely, amplitude-squeezed light should produce the opposite effect (broadening along $\tau$ and sharpening along $z$).

\begin{figure}
\centering{\includegraphics[width=0.45\textwidth]{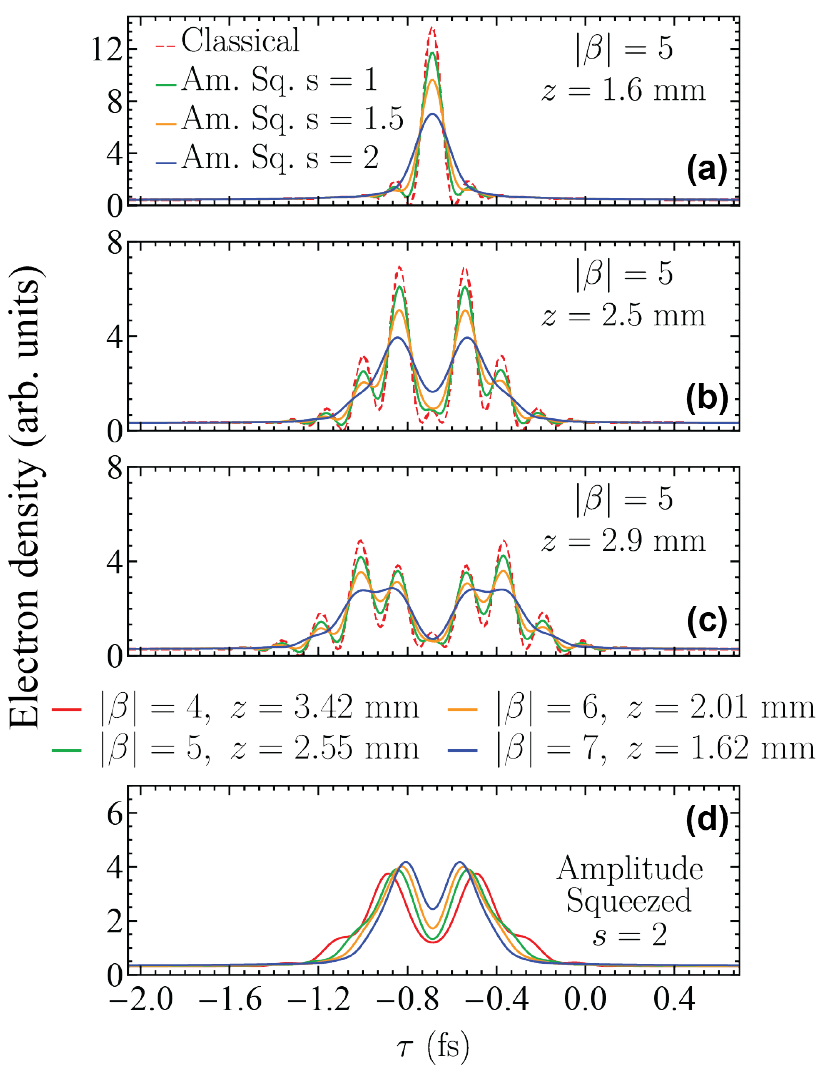}}
\caption{{\bf Tailoring the electron wave packet with amplitude-squeezed light.} (a-c) Electron density profile produced by PINEM interaction with classical (dashed curves) and amplitude-squeezed (solid curves) light after at propagation distance $z$ as indicated by labels. The electron-light coupling coefficient is assumed to be $|\beta|=5$ with $|\beta_0|=0.2$ and $\bar{n}=625$. (d) Evolution of the density profile using amplitude-squeezed light for different coupling strengths $|\beta|$ obtained by varying $|\beta_0|$ with $\bar{n}=625$. We consider $100$\,keV electrons, a photon energy $1.5$\,eV, and a single-mode coupling coefficient $|\beta_0|=0.2$ in all cases.}
\label{Fig3}
\end{figure}

\subsubsection{Synthesis of double-peak electron pulses}

Although PINEM interaction with amplitude-squeezed light renders comparatively poorer focusing, it shows an interesting double-peak pattern for $z$ below the focal spot. This effect, which is already observed in Fig.\ \ref{Fig2}(d), is analyzed in more detail in Fig.\ \ref{Fig3} for different degrees of squeezing. We also show in the same figure the profiles obtained with classical light, revealing amplitude squeezing as a better strategy to produce such double-pulse pattern. We remark that the width and distance between the two pulses can be controlled by varying the coupling strength parameter $|\beta|$ [Fig.\ \ref{Fig3}(d)]. Related to this, we note that a recent experiment \cite{MB20} has shown that a single double-peak electron density profile can be achieved by exploiting classical midinfrared single-cycle laser pulses.

\subsection{Electron compression with minimum-phase-uncertainty light} 

One expects that better focusing can be achieved by reducing phase uncertainty in the optical field. In the limit of large average photon number $\bar{n}\gg1$, the state that produces a minimum phase uncertainty (MPU) has been shown to be given by \cite{KK93}
\begin{align}
\alpha_n\approx\frac{C}{\sqrt{\bar{n}}}{\rm Ai}\left[s_1(1-2n/3\bar{n})\right], \nonumber
\end{align}
where ${\rm Ai}$ is the Airy function, $s_1\approx-2.3381$ is its first zero, $C=\sqrt{2|s_1|/3}/{\rm Ai}'(s_1)\approx2.7805$, and ${\rm Ai}'(s_1)$ is the derivative of ${\rm Ai}$. PINEM focusing with MPU light is illustrated in Fig.\ \ref{Fig2}(b). In contrast to classical light, the Rabi-like oscillations along $z$ are now replaced by a well-defined short-period comb of electron density peaks. This is similar to what we obtain with phase-squeezed light [Fig.\ \ref{Fig2}(c)], but the pattern with MPU light becomes more pronounced. Further deviations from coherent illumination are found in the speed at which compression is achieved: among the statistics under consideration, the shortest FWHM pulse with fixed light intensity and propagation distance is obtained when using MPU light [Fig.\ \ref{Fig2}(e)]. Nevertheless, after a sufficiently large distance $z$, the FWHM reaches similar values with MPU, coherent, and phase-squeezed light, while amplitude-squeezed light systematically leads to lower compression, and this effect becomes more dramatic when increasing the coupling coefficient $|\beta|$ [Fig.\ \ref{Fig2}(f)].

\begin{figure*}
\centering{\includegraphics[width=0.75\textwidth]{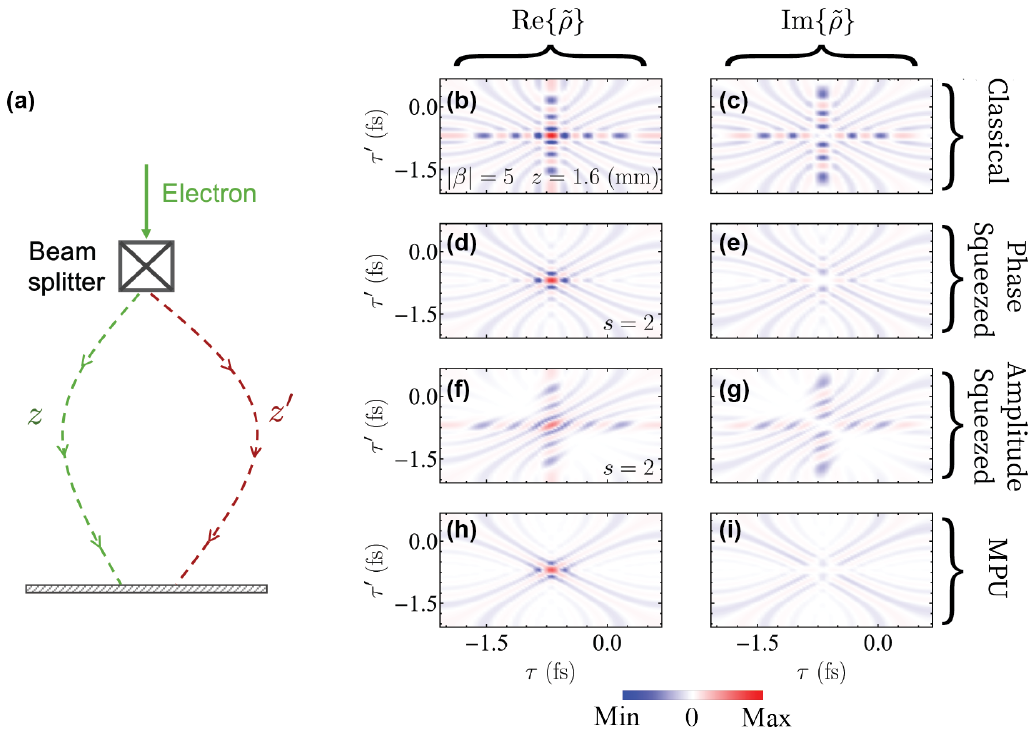}}
\caption{{\bf Measuring the electron density matrix through self-interference}. (a) Sketch of an experimental arrangement to explore electron auto-correlation by means of a beam splitter and different lengths ($z$ and $z'$) along the two electron paths before recombination at the detection region. (b-i) Real (left panels) and imaginary (right panels) parts of the electron density matrix as a function of shifted times $\tau$ and $\tau'$ for $z=1.6$\,mm and different statistics of the PINEM light, as indicated by labels. We consider 100\,keV electrons, 1.5\,eV PINEM photons, a squeezing paramerter $s=2$, and coupling parmameters $|\beta_0|=0.2$ and $|\beta|=5$.}
\label{Fig4}
\end{figure*}

\subsection{Electron self-interference}

We can further modify the focal properties of the electron by mixing it with a delayed version of itself, using for example a beam splitter and different lengths $z$ and $z'$ of the two electron paths converging at the observation region, as sketched in Fig.\ \ref{Fig4}(a). We assume that $z-z'$ is tuned to be a multiple of the electron wavelength, thus rendering $\rho\propto\tilde\rho$ [see Eq.\ (\ref{rhotilde})], considering for simplicity an incident electron plane wave [i.e., $\phi(z-vt)=1/\sqrt{L}$, where $L$ is a quantization length]. Using the notation of Eq.\ (\ref{rho1}), the electron density profile obtained in this way then results from the superposition $(L/2)\sum_n |\psi_n(z,t)+\ee^{\ii\varphi}\psi_n(z',t)|^2=\tilde\rho(z,\tau,\tau)/2+\tilde\rho(z,\tau',\tau')/2+{\rm Re}\{\ee^{-\ii\varphi}\tilde\rho(z,\tau,\tau')\}$, where an overall phase $\varphi$ is introduced (e.g., by means of electrostatic elements along one of the electron arms \cite{VBM18}) to allow us to switch between the real and imaginary parts of $\tilde\rho(z,\tau,\tau')$. An example of how this quantity depends on PINEM light statistics is shown in Fig.\ \ref{Fig4}(b-i), plotted over a discrete dense sampling of $\tau$ and $\tau'$ points satisfying the condition that $v(\tau-\tau')$ are multiples of the electron wavelength. Interestingly, we observe a rotation of the focal spot feature when going from classical to amplitude-squeezed light. This is consistent with the poorer focusing properties observed for the latter. Through the proposed electron self-interference, the focal spot profile can be modified to cover a wide variety of patterns observed for different light statistics. In particular, phase-squeezed and MPU light produce a radical departure in $\tilde{\rho}(z,\tau,\tau')$ relative to classical coherent light.

\begin{figure*}
\centering{\includegraphics[width=0.75\textwidth]{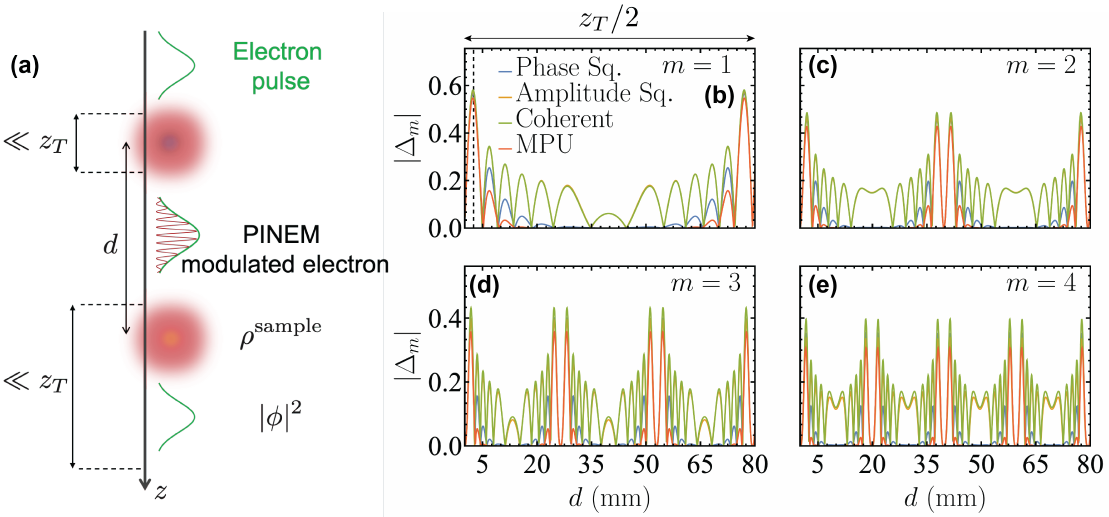}}
\caption{\textbf{Dependence of sample polarization on electron density matrix.} (a) Sketch of an electron wave packed undergoing PINEM modulation, followed by propagation along a distance $d$, and interaction with a single-mode sample of frequency $\omega'_0=m\omega_0$ that is a harmonic $m$ of the PINEM photon frequency. (b-e) Amplitude $\Delta_m$ of the oscillation at frequency $\omega'_0$ displayed by the sample polarization after interaction with the electron. We plot $|\Delta_m|$ for a few values of $m$ as a function of PINEM-sample distance $d$ and different PINEM-light statistics. All parameters are the same as in Fig.\ \ref{Fig4}.}
\label{Fig5}
\end{figure*}

\section{Effect of the electron density matrix on the excitation of a sample}

A commonly asked question relates to how the probability and distribution of excitations produced in a sample are affected by the profile of the beam in an electron microscope. The dependence on the transverse component of the electron wave function has been shown to reduce to a trivial average of the excitation produced by line-like beams over the lateral electron density profile \cite{RH1988,paperarxiv2}. In the present study, we concentrate instead on the longitudinal electron wave function (i.e., along the beam direction). Within first-order Born approximation, the excitation probability is known to be independent of the longitudinal electron wave function when the initial states of the sample and the electron are not phase-correlated \cite{PG19,paperarxiv2}, although a dependence has been shown to arise when the sample state is a coherent superposition of ground and excited states that is phase-locked with respect to the electron arrival time \cite{PG19}, and for example, this effect is actually observed in double-PINEM experiments \cite{PRY17}. Here, we concentrate on the common scenario of a sample prepared in its ground state before interaction with the electron. Remarkably, even when considering higher-order interactions, the number of excitations created by the electron has been shown to still remain independent of the longitudinal wave function \cite{paperarxiv3}, which incidentally implies that the cathodoluminescence intensity is also independent. We generalize this result below by calculating the full density matrix of the bosonic mode, which turns out to have a Poissonian diagonal part equally independent of electron wave function, although the coherences exhibit a dependence on the quantum state of light used in the PINEM interaction to modulate the electron.

For simplicity, we consider a single sample bosonic mode of frequency $\omega'_0$ interacting with an incident PINEM-modulated electron wave packet [Fig.\ \ref{Fig5}(a)]. We can then treat the electron-sample interaction using the same formalism as in Sec.\ \ref{quantumPINEM} by just iterating Eq.\ (\ref{eq:solution}). We find the expression
\begin{align}
&|\Psi(z,t)\rangle=\ee^{\ii k_0z-\ii E_0t/\hbar}\phi(z-vt)\!\!\sum_{\ell=-\infty}^\infty\sum_{n=0}^\infty\sum_{n'=0}^\infty
f_\ell^n f_{-n'}^{\prime n'} \nonumber\\
&\times \ee^{\ii \omega_0[\ell(z/v-t)-n t]-2\pi\ii\ell^2d/z_T-\ii n'\omega'_0z/v} |nn'\rangle \label{lastsupper}
\end{align}
for the wave function of the entire system, comprising the electron, as well as the PINEM and sample bosonic modes, the Fock states of which are labeled by their respective occupation numbers $n$ and $n'$. Primed quantities are reserved here for the sample [i.e., $f_\ell^n$ refers to the first PINEM interaction, while $f_{\ell'}^{\prime n'}$ describes the coupling to the sample in Eq.\ (\ref{lastsupper})], and in particular the condition $\ell'=-n'$ (i.e., sample initially prepared in its ground state $|0\rangle$) is used to write the coefficients $f_{-n'}^{\prime n'}$. Additionally, we introduce a phase correction $\propto\ell^2$ accounting for propagation over a macroscopic distance $d$ separating the PINEM and sample interaction regions, but we neglect this type of correction for relatively short propagation along the extension of the envelope function $\phi(z)$ and within the sample interaction region [see Fig.\ \ref{Fig5}(a)]. The density matrix of the sample mode after interaction with the electron,
\begin{align}
\rho^{\rm sample}=\sum_{n'_1n'_2}\rho^{\rm sample}_{n'_1n'_2} \ee^{-\ii (n'_1-n'_2)\omega'_0t} |n'_1\rangle\langle n'_2|,
\nonumber
\end{align}
is then obtained by tracing out electron (integral over $z$) and PINEM boson (sum over $n$) degrees of freedom. More precisely, we find the coefficients
\begin{align}
&\rho^{\rm sample}_{n'_1n'_2}=\ee^{\ii (n'_1-n'_2)\omega'_0t} \int dz\sum_n\langle nn'_1|\Psi(z,t)\rangle \langle\Psi(z,t)|nn'_2\rangle \nonumber\\
&=f_{-n'_1}^{\prime n'_1} f_{-n'_2}^{\prime n'_2*} \sum_{\ell_1=-\infty}^\infty \sum_{\ell_2=-\infty}^\infty
\phi_{\ell_1\ell_2n'_1n'_2} \sum_{n=0}^\infty f_{\ell_1}^n {f_{\ell_2}^n}^*,
\label{rhonnsum}
\end{align}
where
\begin{align}
\phi_{\ell_1\ell_2n'_1n'_2} = &\ee^{2\pi\ii(\ell_2^2-\ell_1^2)d/z_T} \label{phillnn}\\
&\times \int dz \, |\phi(z)|^2 \, \ee^{\ii[(\ell_1-\ell_2)\omega_0-(n'_1-n'_2)\omega'_0]z/v}.
\nonumber
\end{align}
Incidentally, further electron propagation beyond the sample should also involve corrections to the linearized momentum $n'\omega'_0/v$, on which we are not interested here.

We remind that the momentum decomposition of $\phi$ involves small wave vectors compared with $\omega/v$, so its role in the integral of Eq.\ (\ref{phillnn}) consists in introducing some broadening with respect to the perfect phase-matching condition 
\begin{align}
(\ell_1-\ell_2)\omega_0=(n'_1-n'_2)\omega'_0.
\label{condition}
\end{align}
Such broadening produces nonzero (but small) values of $\phi_{\ell_1\ell_2n'_1n'_2}$ even when $\omega_0/\omega'_0$ is not a rational number. For simplicity, we consider $\omega_0/\omega'_0$ to be a rational number and further assume the spectral width of the sample mode to also be small compared with $\omega_0$; the coefficients of Eq.\ (\ref{phillnn}) then reduce to
\begin{align}
\phi_{\ell_1\ell_2n'_1n'_2}=\ee^{2\pi\ii(\ell_2^2-\ell_1^2)d/z_T}, \nonumber
\end{align}
subject to the condition given by Eq.\ (\ref{condition}).

We note that the diagonal elements $\rho^{\rm sample}_{n'n'}$ involve just $\ell_1=\ell_2$ terms in virtue of Eq.\ (\ref{condition}), so the only nonzero coefficients in Eq.\ (\ref{rhonnsum}) for those elements are $\phi_{\ell\ell n'n'}=1$, and, using the normalization condition $\sum_{\ell n} |f_\ell^n|^2=1$, we find $\rho^{\rm sample}_{n'n'}=|f_{-n'}^{\prime n'}|^2$, which does not depend on the PINEM coefficients $f_\ell^n$: we corroborate that the number of excitations created in the sample is independent of how the incident PINEM electron is prepared \cite{paperarxiv3}; additionally, the distribution of those excitations is also independent. More specifically, upon inspection of Eq.\ (\ref{eq:intcoeff}), we find $f_{-n'}^{\prime n'}=\ee^{\ii \chi'} \ee^{-|\beta'_0|^2/2} {\beta'_0}^{*n'}/\sqrt{n'!}$, and therefore,
\begin{align}
\rho^{\rm sample}_{n'n'}=\left|f_{-n'}^{\prime n'}\right|^2=\ee^{-|\beta'_0|^2}\frac{|\beta'_0|^{2n'}}{n'!}
\nonumber
\end{align}
reduces to a Poissonian distribution regardless of the quantum state of the incident electron, with average $|\beta_0|^2$ corresponding to the contribution of the mode under consideration to the EELS probability. This result, which was found for excitation by an electron treated as a classical probe \cite{SL1971,paper228}, is now generalized to a quantum treatment of the electron. We remark that this conclusion is in essence a result of the nonrecoil approximation.

Combining the above results, the elements of the sample density matrix can be written as
\begin{align}
\rho^{\rm sample}_{n'_1n'_2}=&\ee^{-|\beta'_0|^2} \frac{(-\beta'_0)^{n'_1*}(-\beta'_0)^{n'_2}}{\sqrt{n'_1!n'_2!}} \nonumber\\
&\times {\sum_{\ell_1\ell_2}}'
\ee^{2\pi\ii(\ell_2^2-\ell_1^2)d/z_T} \sum_{n=0}^\infty f_{\ell_1}^n f_{\ell_2}^{n*},
\nonumber
\end{align}
where the sum is subject to the condition imposed by Eq.\ (\ref{condition}). The symmetry property $\rho^{\rm sample}_{n'_1n'_2}=\rho^{{\rm sample}*}_{n'_2n'_1}$ is easily verified from this expression. We can now calculate different observables involving the sample mode, as for example $\propto(a^{\prime\dagger}+a')$. The expectation value of this quantity, which vanishes unless the ratio of sample-to-PINEM mode frequencies $\omega'_0/\omega_0=m$ is an integer, only involves terms in which $n'_1$ and $n'_2$ differ by 1. A straightforward calculation leads to the result
\begin{align}
\langle a^{\prime\dagger}+a'\rangle&=2{\rm Re}\{-\beta'_0\Delta_m\ee^{\ii\omega'_0t}\}, \nonumber
\end{align}
where
\begin{align}
\Delta_m=\ee^{2\pi\ii m^2d/z_T} \sum_{\ell=-\infty}^\infty \ee^{4\pi\ii\ell m d/z_T} \sum_{n=0}^\infty f_\ell^n f_{\ell+m}^{n*}.
\label{Deltam}
\end{align}
This polarization matrix element has been recently shown to exhibit some degree of coherence with the light used to modulate the electron in the first PINEM interaction \cite{paperarxiv3}. We show in Fig.\ \ref{Fig5}(b-e) the dependence of $|\Delta_m|$ on PINEM-sample separation $d$ for a few values of $m$ and different PINEM statistics. This quantity is periodic in $d$ with a period $z_T/2m$, as it is clear from the exponential inside the sum of Eq.\ (\ref{Deltam}). Dramatic differences are observed in $|\Delta_m|$ for different PINEM statistics; in particular, a clear trend is observed toward concentration of $\Delta_m$ at specific distances $d$ when the uncertainty in the light coherence is reduced (i.e., when moving from coherent or amplitude-squeezed light to phase-squeezed light, and eventually to MPU light).

Incidentally, a similar analysis for the $N^{\rm th}$ moment $\propto(a^{\prime\dagger}+a')^N$ leads to a contribution oscillating at frequency $N\omega'_0$ with a coefficient $\Delta_{mN}$. An effect at that order is produced if $mN$ is an integer, a condition that can be met for noninteger values of the sample-PINEM frequency ratio $\omega'_0/\omega_0=m$; for example, an oscillation with frequency $\omega_0$ is induced in $\propto(a^{\prime\dagger}+a')^2$ after electron-sample interaction if the sample mode frequency is half of the PINEM photon frequency.

The time-dependent of the off-diagonal sample density matrix components under discussion could be measured through attosecond streaking \cite{IQY02,SKK07}, as a function of the delay between the times of arrival of the electron and an x-ray pulse, giving rise to oscillations in the energy of photoelectrons produced by the latter as a function of such delay. For low-frequency sample modes, a direct measurement could be based on time-resolved quantum tomography of the sample state; this strategy could benefit from low-frequency beatings resulting from the combination of multiple sample modes of similar frequency. More direct evidence should be provided by the nontrivial interference that has been shown to emerge when mixing the PINEM light with cathodoluminescence emission from the sample \cite{paperarxiv3}.

\section{Conclusions}

We have demonstrated that the interaction of free electrons with quantum light opens a new direction for modulating the longitudinal electron profile, the degree and duration of electron pulse compression, and the statistics associated with this compression. By squeezing the interacting light in phase, the formation of electron pulses is accelerated, and this effect is maximized when using optical fields with an Airy number distribution that minimizes phase uncertainty. Interestingly, amplitude-squeezed light leads to the emergence of double-pulse electron profiles, which could be useful to investigate dynamical processes in a sample. The influence of light statistics becomes more dramatic when examining the electron density matrix after interaction, a quantity that can be accessed through our proposed self-interference experiment. Additionally, we have shown that the excitation of a sample by the electron is affected by how the latter is modulated, and in particular, by the statistics of the modulating light. Indeed, although no dependence is predicted in the probability of exciting sample modes, the temporal evolution of the electron-induced off-diagonal sample density matrix elements shows a dramatic departure from the results observed with laser-modulated electrons when considering instead electrons that have interacted with quantum light. Besides their practical interest to shape and temporally compress free electrons, the results here presented reveal a wealth of fundamental phenomena emerging from the interaction with nonclassical light. We further anticipate potential application in the creation of light sources with nontrivial statistics through electron-induced optical emission using gratings and undulators.

\appendix 
\renewcommand{\thesection}{A} 
\renewcommand{\theequation}{A\arabic{equation}} 


\section{Derivation of Eq.\ (3)}
\label{appendixA}

We review a derivation of Eq.\ (\ref{eq:intcoeff}) presented elsewhere \cite{paper339} to describe the evolution of an electron interacting with a dominant quantized electromagnetic mode in a quantum optics framework \cite{SZ97,M94}. A generalization to multiple modes has also been presented \cite{paper357}. Under the assumptions discussed in Sec.\ 2A, inserting the ansatz solution for the wave function given in Eq.\ (\ref{eq:solution}) into the Schr\"{o}dinger equation $\ii\hbar\partial_t |\psi(\rb,t)\rangle=(\hat{\mathcal{H}}_0+\hat{\mathcal{H}}_{1})|\psi(\rb,t)\rangle$ for the Hamiltonian defined in Eqs.\ (\ref{H0H1}), we find the differential equation
\begin{align}
    \partial_z f_{\ell}^{n}=\sqrt{n} \, u^* \, f_{\ell+1}^{n-1} -\sqrt{n+1}\,u\,f_{\ell-1}^{n+1},\label{eq:diffequation}
\end{align}
for the expansion coefficients, where $u_j(z)=(e/\hbar\omega_0)\mathcal{E}_{0,z}(z)\ee^{-\ii\omega_0z/v}$. We note that Eq.\ (\ref{eq:diffequation}) preserves the sum $n+\ell$, thus guaranteeing that the number of excitations in the electron-boson system is conserved along its evolution. This property ensures that the problem can be mapped onto a classically driven quantum harmonic oscillator (QHO), which admits an analytical solution \cite{G1963}. The connection between the two systems is made clear by writing the Hamiltonian $\hat{\mathcal{H}}=\hbar \omega a^\dagger a+g(t)a+g^*(t)a^\dagger$ for the QHO, along with its wave function $|\psi(t)\rangle=\sum_{n}c_{n}(t)\ee^{-\ii n\omega t}|n\rangle$, whose coefficients follow the  equation of motion $\ii\hbar  \partial_t c_{n}=\left[\sqrt{n}\,g^*\,c_{n-1}\ee^{\ii\omega t}+\ee^{-\ii\omega t}\sqrt{n+1}\,g\,c_{n+1}\right]$.
For an initial state written as $|\psi^{\rm I}(t_0)\rangle=\sum_{n}c_{n}(t_0)|n\rangle$ in the interaction picture, the solution at later times can be expressed as s $\langle n |\psi^{\rm I}(t)\rangle = \langle  n|\hat{\mathcal{S}}(t,t_0)|\psi^{\rm I}(t_0)\rangle$ in terms of the scattering operator \cite{G1963}
\begin{align}
    \hat{\mathcal{S}}(t,t_0)  = \ee^{\ii \chi} \ee^{\beta_0^* a^\dagger-\beta_0 a },\label{eq:smatrix}
\end{align}
where $\beta_0(t,t_0)=\frac{\ii}{\hbar}\int_{t_0}^{t}dt'g(t')\ee^{-\ii \omega t'}$ is the coupling coefficient and $\chi = -\frac{1}{\hbar}\int_{t_0}^{t}dt' {\rm Re}\{\beta_0(t',t_0)g^*(t')\ee^{\ii\omega t'}\}$ is a global phase; we obtain $c_n(t)=\sum_{m=0}^\infty c_{m}(t_0)   \langle n |  \hat{\mathcal{S}}(t,t_0) | m\rangle$. Then, an explicit expression for the matrix elements of the evolution operator [Eq.\ (\ref{eq:smatrix})] can be obtained by applying the Baker-Campbell-Hausdorff formula:
\begin{align}
     \langle n |  \hat{\mathcal{S}}(t,t_0) | m\rangle=& \sqrt{m!n!}~\ee^{-|\beta_0|^2/2}(-\beta_0)^{m-n} \nonumber \\
    &\times \!\!\!\!\!\! \sum_{n'=\max\{0,n-m\}}^{n} \! \frac{(-|\beta_0|^2)^{n'}}{n'!(m-n+n')!(n-n')!}. \nonumber
\end{align}
We now connect these results with the solution of our electron-boson system by exploiting the mapping $f^n_{n_0-n}=c_n$ enabled by the conservation of $n_0=n+\ell$. Making the substitutions 
$g\ee^{-\ii\omega t}\rightarrow -\ii \hbar v u$ and $t\rightarrow z/v$, and imposing the condition $f_\ell^n(z\rightarrow -\infty )=\delta_{\ell,0}\alpha_n$ to Eq.\ (\ref{eq:diffequation}), where $\alpha_m$ are the initial coefficients of the initial quantum light state $\sum_n\alpha_n|n\rangle$, we readily find Eq.\ (\ref{eq:intcoeff}).

\section{Derivation of Eq.\ (10)}
\label{appendixB}

Starting from Eq.\ (\ref{eq:intcoeff}) and considering $n\gg1$ and $|\ell|\ll n$ for typical single-mode coupling conditions $|\beta_0|\ll1$, the dominant contribution to the sum comes from $n'\ll n$ terms, so we can approximate $F_\ell^n\approx\sum_{n'=0}^\infty\frac{(-1)^{n'}|\beta_0|^{2n'+\ell}}{n'!(\ell+n')!} \frac{\sqrt{(n+\ell)!n!}}{(n-n')!}$. We now apply the Stiling formula $n!\approx\sqrt{2\pi n}\,(n/e)^n$ to the factorials in the rightmost fraction and neglect $\ell$ and $n'$ in front of $n$ in the factors that are not affected by an exponent $n$. This allows us to approximate $\sqrt{(n+\ell)!n!}/(n-n')!\approx (n/e)^{n'+\ell/2}\ee^M$, where $M=n\,\ln\left[\sqrt{1+\ell/n}/(1-n'/n)\right]$. We then retain only terms up to first order in the Taylor expansion of the logarithm to find $M\approx n'+\ell/2$. Upon insertion of this result into the above expression for $F_\ell^n$, we find $F_\ell^n\approx\sum_{n'=0}^\infty(-1)^{n'}|\sqrt{n}\beta_0|^{2n'+\ell}/[n'!(\ell+n')!]$, which directly yields Eq.\ (\ref{largen}) by identifying the sum as the Taylor expansion of the Bessel function $J_n$ with argument $2\sqrt{n}|\beta_0|$.



\section*{Acknowledgments} 
\noindent We thank Ofer Kfir and Claus Ropers for many helpful and enjoyable discussions. ERC (Advanced Grant 789104-eNANO), Spanish MINECO (MAT2017-88492-R and SEV2015-0522), the Catalan CERCA Program, and Fundaci\'{o} Privada Cellex. V.D.G. acknowledges support from the EU (Sk\l{}odowska-Curie Grant 713729).


%


\end{document}